\documentclass[10pt,journal,compsoc]{IEEEtran}
\IEEEoverridecommandlockouts
\usepackage{cite}
\usepackage{amsmath,amssymb,amsfonts}
\usepackage{algorithmic}
\usepackage{graphicx}
\usepackage{textcomp}
\usepackage{bm}
\usepackage{makecell}
\usepackage{hhline}
\usepackage{caption}
\usepackage[table]{xcolor} 
\usepackage{booktabs} 
\usepackage{hyperref}
 \usepackage{balance} 
\definecolor{gray}{rgb}{0.35, 0.35, 0.35}

\hypersetup{
    colorlinks=true,   
    linkcolor=blue,    
    citecolor=blue,    
    urlcolor=blue,      
}

\def\BibTeX{{\rm B\kern-.05em{\sc i\kern-.025em b}\kern-.08em
		T\kern-.1667em\lower.7ex\hbox{E}\kern-.125emX}}

\usepackage{graphicx}
\usepackage{booktabs}
\usepackage{multirow}
\usepackage{subcaption}
\usepackage{tabularx}
\usepackage[ruled,linesnumbered]{algorithm2e} 
\usepackage{pifont}

\def\bf{\boldsymbol{f}}

\definecolor{frenchblue}{rgb}{0.0, 0.45, 0.73}

\usepackage{color}

\begin{document}
	
	\title{
	\textcolor{black}{From Signals to Evidence: Interpretable Underwater Acoustic Target Sensing\\ with Large Language Models \\— Part II: Passive Detection}
	}

    \author{
    \thanks{\rm \textcolor{black}{The code is released at \url{https://github.com/xilyao/CFD-DC}.}}
    }
	

		\author{Zhiyuan Li, Yuge Zhang, Jie Chen,  Xiruo Su, Zhengqiao Zhao, Xiling Yao \\Lingji Xu, Jingdong Chen, Yixin Yang
        \IEEEcompsocitemizethanks{
		\IEEEcompsocthanksitem Corresponding author: Jie Chen.
    \IEEEcompsocthanksitem   

     Z. Li, Y. Zhang, J. Chen, Z. Zhao, X. Su, X. Yao, J.-D. Chen are with School of Artificial Intelligence, Northwestern Polytechnical University, Xi’an, Shaanxi 710129, China (emails:      lzynwpu@mail.nwpu.edu.cn, zhangyuge@mail.nwpu.edu.cn, dr.jie.chen@ieee.org, xiruosu@nwpu.edu.cn, zhengqiao.zhao@nwpu.edu.cn, xilingyao@mail.nwpu.edu.cn, jchen@nwpu.edu.cn).

     L. Xu is with School of Ocean Engineering and Technology, Sun Yat-sen University, Zhuhai, 519000, Guangdong, China (xulj26@mail.sysu.edu.cn).

    Y. Yang is with School of Marine Science and Technology, Northwestern Polytechnical University, Xi’an, Shaanxi 710129, China, and also with School of Electronic and Optical Engineering, Nanjing University of Science and Technology, Nanjing, Jiangsu 210094, China (email: yxyang@nwpu.edu.cn).
    }
    }

	\IEEEtitleabstractindextext{
	\begin{abstract}
Underwater acoustic passive detection is essential to maritime surveillance and intelligent ocean sensing. Existing detection systems typically collapse multichannel array observations into a binary decision or a scalar confidence score, offering limited insight into the physical evidence that supports or contradicts a target declaration. This limitation is acute at low signal-to-noise ratio (SNR), where isolated spectral peaks and transient beamforming responses are inherently ambiguous and physically verifiable targets are characterized by an ensemble of coupled reinforcing cues. Building on the unified from-signals-to-evidence framework introduced in the companion Part~I~\cite{zhang2026part1}, this paper instantiates the framework for passive target detection. A physics-aware multichannel acoustic encoder encodes the array observations into complementary acoustic evidence tokens through a direction-guided architecture that combines minimum variance distortionless response (MVDR) processing with time--frequency and fine-frequency representations. These tokens are combined with environmental priors and fed into a pretrained large language model (LLM) that evaluates target-present and noise-only hypotheses and generates a structured detection report detailing target presence, bearing, spectral-line locations, received SNR, range, and per-field confidence scores. A physics-parameterized data engine extended to realistic oceanic channels provides traceable supervision at both discrete-label and structured-description levels. Experiments under white-noise and mixed ocean-background conditions demonstrate reliable detection, accurate bearing estimation, and physically interpretable structured reports, validating the feasibility of evidence-grounded underwater acoustic detection.

	\end{abstract}
	
	}
	\maketitle
	\IEEEdisplaynontitleabstractindextext
	
	\IEEEpeerreviewmaketitle
	
	\section{Introduction}
	\label{sec:intro}
	
Underwater acoustic passive detection (UAPD), which aims to concurrently detect and characterize targets from multichannel array observations, is fundamental to maritime surveillance, ecological monitoring, and ocean exploration~\cite{Jin2023OffGridDE,yu2025hierarchical,zhang2025unsupervised}. In passive settings, target perception relies heavily on spatial and spectral signatures embedded within the received acoustic wavefield, typically exposed via time-frequency analysis, spatial localization, and the extraction of machinery-related spectral components. However, these signatures are routinely masked and distorted by ambient ocean noise, multipath reverberation, and competing acoustic interferers, severely degrading spatial coherence and obscuring critical acoustic signatures~\cite{tong2024adaptive,su2025underwater}. Consequently, robustly suppressing interference, resolving source bearings, and isolating reliable spectral features remain challenges.

\begin{figure*}[!t]
	\centering
	\includegraphics[width=1\textwidth]{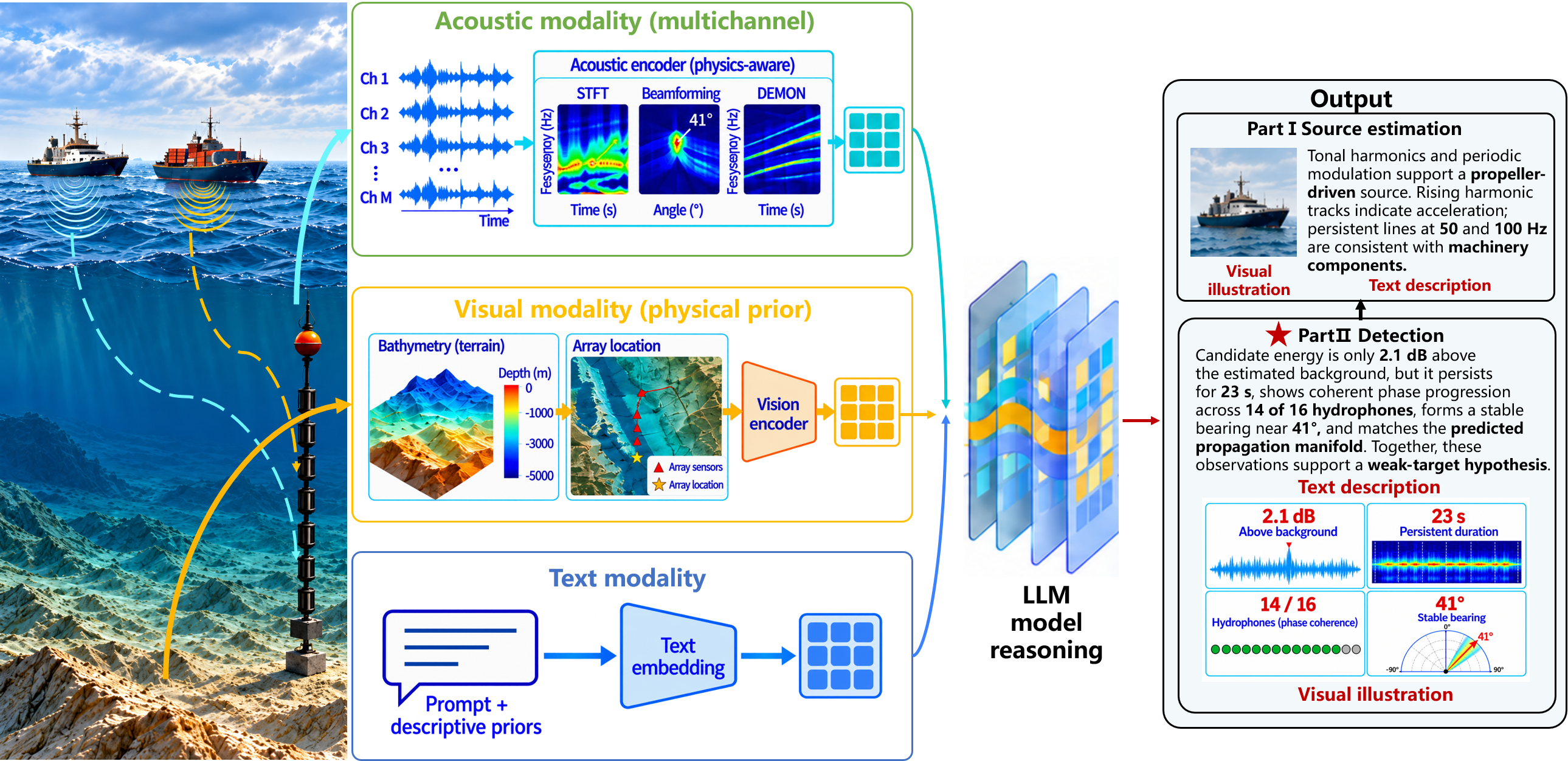}
	\caption{Conceptual overview of the proposed UPD framework for interpretable underwater acoustic target detection. \emph{Part I: Framework \& Recognition} has been addressed in our previous work\cite{zhang2026part1}.}
	\label{fig:overall}
\end{figure*}

The methodology for underwater target detection has steadily evolved from classical statistical hypothesis testing toward data-driven learning. Conventional detectors typically formulate statistics based on energy integration, matched filtering, spectral line detection, beamformed outputs, or spatial coherence metrics, comparing this statistic against an analytical threshold~\cite{caldeira2023eemd,Pan2021TimefrequencyDC}. While theoretically grounded in statistical physics, these methods suffer from pronounced performance degradation when real-world oceanic conditions deviate from idealized assumptions regarding noise stationarity, source dynamics, or propagation channels. More recently, deep neural networks (DNNs) have improved detection performance under low signal-to-noise ratio (SNR) conditions by extracting discriminative representations directly from raw waveforms, spectrograms, spatial spectra, or raw multichannel tensors~\cite{taherian2020robust,liu2020multichannel,trucco2021supervised}. Nevertheless, the prevailing decision paradigm remains intrinsically reductionist: spatially continuous acoustic fields are collapsed into a single scalar confidence score or a binary indicator. Such scores indicate model confidence but fundamentally fail to  articulate which observations corroborate the hypothesis, or explain why a subtle signature represents an authentic physical target rather than a stochastic background fluctuation.

This limitation becomes critical in the low-SNR regime. An isolated spectral peak is inherently ambiguous, as ambient noise and transient interference generate false spectral structures. Similarly, transient beamforming peaks often stem from coherent background interference or acoustic environmental mismatch~\cite{fan2023fast,Song2023TensorBasedSB}. Conversely, a physically verifiable target is characterized by an ensemble of coupled reinforcing physical cues: tonals persist across temporal frames, harmonic series and modulation sidebands evolve coherently, inter-channel phase differences align with admissible directions, and spatial response patterns obey array manifold constraints and environmental propagation limits~\cite{Wang2024ADE,Li2022DeepLD,su2025snr}. Therefore, detection should be reformulated as a process of multi-evidence hypothesis reasoning.

Recent advances in large language models (LLMs) and multimodal foundation models present an opportunity to operationalize this reasoning paradigm. The transformer architecture excels at capturing long-range contextual dependencies, enabling the aggregation of weak, temporally persistent signals submerged within ambient fluctuations~\cite{Gao2024PETransformer,Chen2023HTDet,Ji2024FBDPN}. Simultaneously, multimodal cross-attention mechanisms allow acoustic wavefields to be conditioned on explicit domain priors, including array geometry, sound speed profiles (SSPs), bathymetry, and acoustic transmission loss ~\cite{Sun2025HybridDOT,Tan2026TargetStrength,Li2026FourDSSF}. The autoregressive generation capability of LLMs also enables systems to corroborate evidence and uncertainty estimates. However, directly applying generic LLMs to underwater acoustic detection presents key limitations. Most conventional audio backbones are designed for human auditory perception and dominated by prominent, high-energy patterns, making them insensitive to the weak, phase-coherent signals essential for sonar detection. In addition, standard audio tokenization overlooks array-domain physics, discarding inter-sensor spatial coherence and critical array processing gains. Consequently, forcing LLMs to reason over abstract latent embeddings without physical meaning leads to unreliability of detection decisions.

To overcome these barriers, this work extends the Underwater acoustic Physics-aware Description (UPD) framework from \emph{Part I: Recognition}~\cite{zhang2026part1} to \emph{Part II: Passive detection}. Our framework discretizes the multichannel array wavefield into a structured set of complementary physical evidence tokens, each representing a verifiable attribute of candidate targets. These tokens preserve the spatio-temporal principles of the acoustic field, enabling candidate targets to be evaluated across temporal persistence, spatial distribution, propagation viability, and background noise context. To anchor this reasoning process in real-world ocean environments, the acoustic tokens are cross-conditioned with non-acoustic multimodal priors, including contextual prompts and environmental parameters (e.g., sea state, SSPs, array geometries, and bathymetric grids). The aligned LLM functions as an evidence-grounded reasoning engine. Conditioned on both physical tokens and environmental priors, the model evaluates target-present and noise-only hypotheses to generate a structured detection report. This output provides an interpretable analysis that details target motion, bearing tracks, discrete machinery tonals, and explicit confidence bounds. Furthermore, we extend our physics-based acoustic simulation pipeline to realistic oceanic channels, providing discrete spatial-spectral labels and structured diagnostic texts, which demonstrate the availability of UPD in real underwater cases.

This work is organized in two parts. The two parts are organized by the maturity of signal conditions rather than by the order of the processing chain. 

\emph{Part I}, establishes the general framework and validates it on the recognition task, where signals have undergone front-end preprocessing and the physical priors are comparatively well defined, offering the most direct demonstration of the proposed paradigm. 

\emph{Part II}, the present paper, advances to target detection under passive conditions, which operates closer to the raw multichannel measurements and demands a deeper integration of acoustic physics into the framework.

The main contributions of this paper are summarized as follows:
\begin{itemize}
\item We adapt evidence-grounded underwater acoustic understanding from target recognition (Part I) to passive detection and formulate detection as physics-grounded multi-evidence hypothesis reasoning, in which a target declaration is supported by explicit time–frequency, temporal, spatial, propagation, and background evidence.

\item We develop a physics-aware multichannel acoustic encoder that preserves weak target structures across the time, frequency, and array dimensions and converts them into physically meaningful evidence tokens, including spectral persistence, temporal continuity, directional concentration, inter-channel coherence, array-manifold consistency, and environmental background descriptors.

\item We develop an LLM-based evidence reasoning mechanism that jointly analyzes supporting and conflicting observations and generates a structured detection description containing target presence, direction-of-arrival information, candidate spectral components, temporal behavior, spatial and propagation consistency, environmental conditions, and uncertainty.

\item Extensive experiments demonstrate reliable target detection through joint representation of complementary physical cues and multimodal reasoning, with consistent associations among spectral structures, directional responses, temporal persistence, and propagation priors, validating the feasibility of evidence-grounded underwater acoustic detection.
\end{itemize}




The remainder of this paper is organized as follows. Section 2 reviews related work. Section 3 presents the UPD framework. Section 4 describes the physics-based data construction. Section 5 reports experimental results, followed by discussion in Section 6 and conclusions in Section 7.

   

\section{Related Work}

\subsection{Underwater Passive Detection}

The detection of underwater acoustic arrays aims to infer the existence, spatial location, physical attributes, and temporal evolution of acoustic sources from multichannel measurements. Classical detection primarily relies on spatial coherence and array manifolds. Conventional beamforming (CBF), along with high-resolution methods such as the minimum variance distortionless response (MVDR) beamformer, multiple signal classification (MUSIC), and spatial-spectrum reconstruction based on deconvolution~\cite{Chen2015EffectsOS,meng2023wind,Doclo2003DesignOF,Yang2020DeconvolutionOD}, have been widely used in real sea trials. While background noise and reverberation can distort these spatial relationships and consequently degrade the performance of the DOA estimation~\cite{tong2024adaptive}. 

The past decade has witnessed increasing use of neural networks for underwater DOA estimation and source localization~\cite{Sun2025HybridDOT,Liu2026ArrayLocalization,SU2BYAA}. These approaches aim to alleviate the dependence of conventional methods on accurate covariance estimation and exhaustive spatial-spectrum searches. For example, Xu et al. proposed a sparsely gated mixture-of-experts network under hybrid noise conditions~\cite{Xu2026SMoEDOA, Huang2025MoELocNet}. Gao et al. integrated a conditional diffusion model with multi-output Gaussian processes for continuous forecasting of range-dependent ocean sound-speed fields~\cite{Gao2026SSFForecast}. For range- and depth-dependent localization, matched-field processing (MFP) has been combined with neural networks~\cite{Song2024LFDMFP,Wang2025MNFFSBL,Qu2024CNNTransformer,Liu2023BCCNN}. Park et al.~\cite{Park2025PINNMFP} employed a physics-informed network to predict replica fields from sparse acoustic-pressure observations and known sound-speed profiles, thereby enabling matched-field range estimation without requiring detailed seabed geoacoustic parameters. More recently, Li et al. combined frequency-difference processing with a multi-branch multitask network to jointly estimate source range and depth under sound-speed-profile mismatch~\cite{Li2026FrequencyDifferenceMTL,Li2025FewShotRange}. 

Temporal sensing further extends these instantaneous estimates to long-term observations and dynamic target behavior. Continuous analysis of bearing and frequency evolution can help distinguish persistent target-related components from random noise and transient interference~\cite{Ali2024NeuralTimeSeries,Liang2024PassiveToneDetection,Zhang2025VariableFrequencyTBD}. Su et al. proposed a method that exploits the context continuity of spectra across multiple frames to identify frequency-bearing bins\cite{su2026underwater}. More generally, once source direction, Doppler shift, or other motion-related quantities have been estimated,  Bayesian state-space models can be applied to reconstruct continuous target trajectories and motion states. Sun et al.~\cite{Sun2023JEAUKF}, for example, incorporated the unknown source center frequency and turn rate into a joint vector and developed an unscented Kalman filter to estimate the position and motion state of an underwater target.  

Related work on acoustic sensing and cooperative classification also explores how complementary observations can be combined through efficient processing and learned feature exchange~\cite{yang2026distributedacoustic,yao2026cfddc}. Nevertheless, existing underwater acoustic sensing systems often remain fragmented across individual tasks. Detection, DOA estimation, ranging, environmental characterization, and tracking are commonly addressed using separate models and task-specific representations. Such fragmentation becomes increasingly restrictive when acoustic sensing is expected to support more general scene understanding.

\subsection{Large Audio-Language Models for Acoustic Sensing}
Large audio-language models (LALMs) extend the instruction-following and text-generation capabilities of pretrained language models to acoustic observations~\cite{tang2024salmonn,chu2024qwen2audio,goel2025audioflamingo3}. Representative systems include SALMONN~\cite{tang2024salmonn}, Qwen2-Audio~\cite{chu2024qwen2audio}, and Audio Flamingo 3~\cite{goel2025audioflamingo3}. Their interfaces convert acoustic features into embeddings that the LLM processes alongside textual instructions, enabling question answering and description generation from audio observations. Feature projection maps encoder outputs into the language-model input space, while query-based alignment also aggregates information across feature sequences. For example, BLIP-2~\cite{li2023blip2} introduces a Q-Former to bridge frozen image encoders and language models. SALMONN uses a window-level Q-Former to combine complementary speech and general-audio features into audio tokens for its LLM. These developments enable specialized acoustic representations to be incorporated into language models, with the encoder and alignment module jointly shaping the information available for subsequent prediction and description.

Spatial audio-language research adapts this framework to sound events and their physical attributes. Zheng et al. propose BAT~\cite{zheng2024bat}, which couples a binaural spatial audio encoder with LLaMA-2 to answer questions about sound event, source directions and distances, and spatial relationships. Tang et al.~\cite{tang2024spatial} combine speech features and spatial intensity vectors through a Q-Former, enabling a Vicuna LLM to generate source direction estimates. ACORN~\cite{wang2025acorn} connects magnitude and phase representations to Qwen and Llama models for direction and range estimation, Doppler analysis, and propagation-related questions. More recently, SAIL~\cite{luo2026sail} represents acoustic content and spatial information in separate streams and connects them to an LLM through a dual-stream Q-Former with queries organized by source. These studies suggest that acoustic encoders, task supervision, and cross-modal interfaces can work together to support descriptions of what is present, where it is located, and how the observed sound relates to its physical surroundings.

The spatial and physical audio-language studies above establish relevant modeling approaches for acoustic sensing. The companion Part I~\cite{zhang2026part1} connects physics-aware acoustic representations to language descriptions for underwater acoustic recognition. The present study addresses the upstream task of target detection and parameter estimation. A central question is how spatial, spectral, and statistical information can be represented together and conveyed through language generation while preserving physical meaning and numerical accuracy. This places acoustic--language alignment within the complete sensing process, from the received observation to a account of target presence, location, tonal characteristics, and signal quality.

\section{From-Signals-to-Evidence Framework: A Brief Recall}
\label{sec:framework}

\begin{figure*}[!t]
	\centering
	\includegraphics[width=0.95\textwidth]{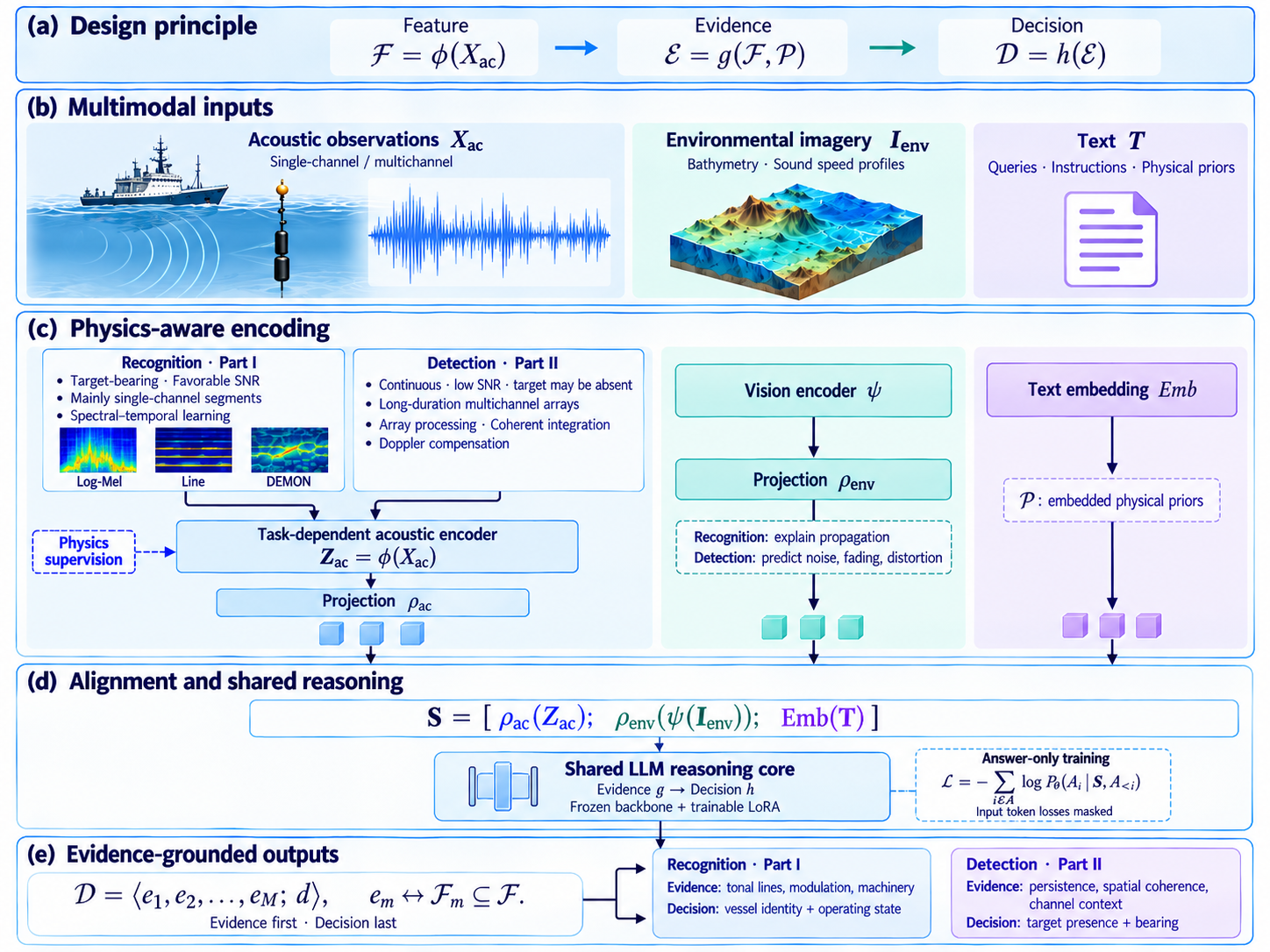}
	\caption{Architecture of the proposed from-signals-to-evidence framework.}
	\label{fig:schema}
\end{figure*}

This section recalls the UPD framework established in Part~I~\cite{zhang2026part1} and highlights the adaptations required for passive detection. The framework is depicted in Fig.~\ref{fig:schema}.

\subsection{Feature--Evidence--Decision Chain}
\label{sec:philosophy}

The UPD framework organizes underwater acoustic sensing around an ordered \emph{feature--evidence--decision} chain,
\begin{equation}
\mathcal{F} = \phi(X_{\mathrm{ac}}), \qquad
\mathcal{E} = g(\mathcal{F}, \mathcal{P}), \qquad
\mathcal{D} = h(\mathcal{E}),
\label{eq:chain}
\end{equation}
where $X_{\mathrm{ac}}$ denotes acoustic observations, $\mathcal{P}$ the embedded textual physical priors, $\phi$ the physics-aware acoustic encoder, and $g$, $h$ the evidence-forming and decision-making processes realized by a shared LLM reasoning core. The decision $\mathcal{D}$ is conditioned on the stated evidence $\mathcal{E}$ rather than directly on the features $\mathcal{F}$, distinguishing this chain from post-hoc explanation schemes. Two principles govern the design: \emph{physical grounding at every stage}, ensuring that features correspond to identifiable acoustic structures rather than opaque activations, and \emph{a shared semantic space} in which acoustic signals, environmental imagery, and textual priors are aligned into the LLM embedding space.

\subsection{Multimodal Input Layer}
\label{sec:input}

The input layer accommodates heterogeneous information without architectural modification: (i) \emph{acoustic modality}---waveform-level observations from single or multiple hydrophones, possibly preprocessed by beamforming; (ii) \emph{environmental modality}---sound speed profiles, bathymetric charts, sea state, and seabed properties, expressed as imagery or structured parameters; and (iii) \emph{textual modality}---physical priors and expert knowledge in natural language. New sources are incorporated by adding encoders or extending the textual context.

\subsection{LLM Reasoning Core and Evidence-Grounded Output}
\label{sec:core}

All modalities are organized into a unified token sequence,
\begin{equation}
\mathbf{S} = \big[\, \rho_{ac}(\mathbf{Z}_{\mathrm{ac}});\ 
\rho_{env}(\psi(\mathbf{I}_{\mathrm{env}}));\ 
\mathrm{Emb}(\mathbf{T}) \,\big],
\label{eq:sequence}
\end{equation}
where $\mathbf{Z}_{\mathrm{ac}}=\phi(X_{\mathrm{ac}})$ denotes acoustic tokens, $\rho_{ac}$ and $\rho_{env}$ are modality-specific projections, $\psi$ encodes environmental imagery, and $\mathrm{Emb}(\mathbf{T})$ embeds textual tokens. The LLM processes $\mathbf{S}$ and generates responses in an ordered evidence-then-decision manner. Pretrained parameters are frozen and adapted through low-rank modules; training supervises only the answer portion of each sequence,
\begin{equation}
\mathcal{L} = - \sum_{i \in \mathcal{A}} 
\log P_{\theta}\!\left( A_i \,\middle|\, \mathbf{S}, A_{<i} \right),
\label{eq:loss}
\end{equation}
where $\mathcal{A}$ denotes answer token positions. The output is a structured description,
\begin{equation}
\mathcal{D} = \langle\ e_1, e_2, \dots, e_M;\ d \rangle, 
\qquad e_m \leftrightarrow \mathcal{F}_m \subseteq \mathcal{F},
\label{eq:output}
\end{equation}
in which each evidence item $e_m$ is linked to a verifiable subset $\mathcal{F}_m$ of the acoustic features, and a fixed field order enforces evidence-before-decision generation.

\subsection{From Recognition to Detection: What Changes}
\label{sec:unified}

Detection and recognition are two instantiations of the same chain, sharing a common semantic core but differing in four key respects~\cite{zhang2026part1}. (i) Operating conditions: recognition typically operates on target-bearing segments with relatively favorable SNR and well-defined physical priors, whereas detection must continuously process raw observations in which a target may be deeply buried in noise. (ii) Nature of input: recognition primarily uses single-channel, target-centered recordings, whereas detection processes multichannel array signals over extended durations and provides spatial information, such as target bearing, in addition to target-presence decisions. (iii) Physical front-end integration: detection relies more heavily on parameterized physical processing, including array processing, coherent integration, and Doppler compensation, incorporated before or within the encoder. (iv) Role of the environment: in recognition, environmental priors help explain how propagation shapes observed acoustic features; in detection, environmental modeling must also anticipate the effects of the noise field, fading, and channel-induced distortion on potential target signatures, thereby supporting reliable detection under adverse acoustic conditions.



\begin{figure*}[!t]
\centering
\includegraphics[width=\textwidth]{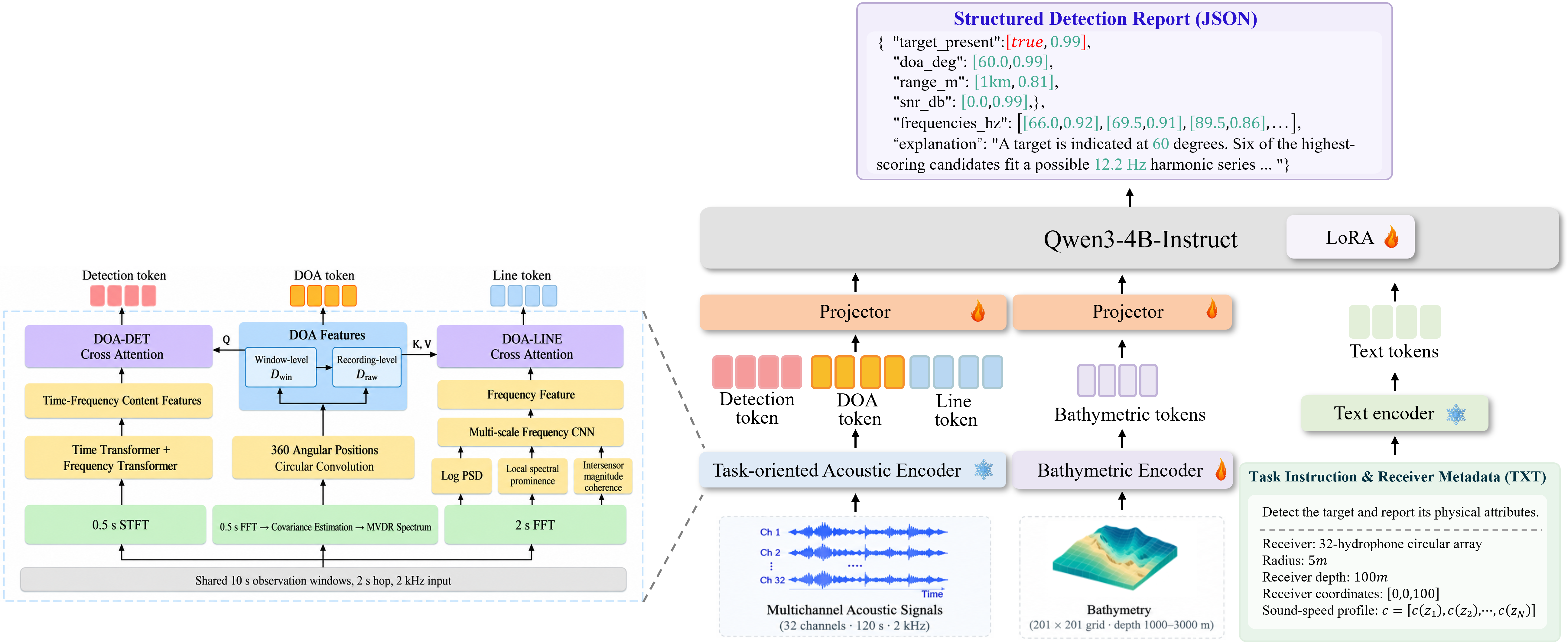}
\caption{Task-oriented acoustic encoding and language alignment. The DOA, DET, and LINE branches share the same observation windows. Their internal Fourier analysis windows determine the spectral grids. DOA provides window-level directional queries to DET and recording-level directional keys and values to LINE. The three sequences and task summaries are projected into the language-model input space. A separate head reads frequency scores from the additional readout position.}
\label{fig:structured_output}
\end{figure*}

\section{Detection Instantiation: Physics-Guided Acoustic Encoding}
\label{sec:detection_instantiation}

UPD estimates target presence, bearing, and spectral-line locations from synchronized array recordings. Its branches are named by their tasks: bearing estimation (DOA), target detection (DET), and spectral-line detection (LINE). Directional features from DOA guide feature aggregation in DET and LINE. The resulting acoustic representations and environmental inputs are aligned with a language model to produce structured reports.

\subsection{Physical Principles of Passive Detection}
\label{sec:low_snr_principles}

Passive detection must distinguish weak target structure from background fluctuations while the source bearing and spectral lines are unknown. A local energy peak at an individual sensor provides limited evidence. The array and the observation duration provide additional information through spatial relationships and temporal and spectral structure.

Let \(M_{\mathrm a}\) be the number of sensors, \(f\) the frequency, and \(\theta_0\) the bearing of a locally planar target component. Let \(\bm x(f)\in\mathbb C^{M_{\mathrm a}}\) be the array spectrum, \(s(f)\) the target's complex spectral amplitude, and \(\bm n(f)\in\mathbb C^{M_{\mathrm a}}\) the background. The steering vector \(\bm a(f,\theta_0)\) depends on relative sensor positions, sound speed, and bearing. Under a narrowband approximation,
\begin{equation}
  \bm{x}(f)=\bm a(f,\theta_0)s(f)+\bm n(f).
  \label{eq:array_model}
\end{equation}
The target-absent case has \(s(f)=0\). When a target is present, its propagation delays produce phase relationships across sensors. Spatial covariance retains these relationships, and a directional spectrum evaluates candidate bearings. The full angular profile supplies directional information for aggregating acoustic features.

A tonal component that is approximately stationary within an analysis window concentrates energy near its frequency. Short Fourier analysis windows retain local time-frequency variation; longer windows provide a finer spectral grid. Successive observation windows provide repeated views of the signal. Their ordered features retain temporal changes, while the means and standard deviations of fine-spectral features summarize average strength and variability. Directional information guides the aggregation of these complementary features for detection and line prediction.

\subsection{Direction-Guided Acoustic Encoding}
\label{sec:direction_guided_encoding}

Let \(B\) be the batch size and \(N\) the number of samples per sensor. The synchronized input is \(\mathbf X\in\mathbb R^{B\times M_{\mathrm a}\times N}\). Each recording lasts 120~s and contains 32 channels sampled at 2~kHz. All three branches share 10~s observation windows with a 2~s hop, giving \(K=56\) windows per recording.

Within each observation window, DOA and DET use 0.5~s Fourier analysis windows, whereas LINE uses 2~s analysis windows. At the common sampling rate, the corresponding transforms contain 1{,}000 and 4{,}000 points, with frequency-bin spacings of 2 and 0.5~Hz. Table~\ref{tab:fft} gives the analysis hops and frequency bands.

Let \(N_{\theta}\) and \(N_f\) denote the numbers of bearing-grid and fine-frequency-grid positions. The internal and output feature widths are \(d\) and \(d_{\mathrm o}\), respectively. Figure~\ref{fig:structured_output} shows the shared windowing and the directional connections between branches.

\subsubsection{Bearing estimation branch DOA}
\label{sec:direction_branch}

The DOA branch estimates spatial covariance within each observation window and computes the minimum variance distortionless response (MVDR) spectrum. Let \(J\) be the number of spectral snapshots per window and \(\bm x_{k,j}(f)\in\mathbb C^{M_{\mathrm a}}\) the array spectrum at frequency \(f\) in snapshot \(j\) of window \(k\). For one observation, the covariance estimate is \(\bm R_k(f)\in\mathbb C^{M_{\mathrm a}\times M_{\mathrm a}}\). Its diagonally loaded form \(\widetilde{\bm R}_k(f)\) adds a positive multiple of the mean diagonal power to every diagonal entry. Let \(P_k(f,\theta)\) denote the MVDR directional power at candidate bearing \(\theta\), and let \(H\) denote conjugate transpose. Then~\cite{uad_capon1969}
\begin{equation}
\begin{aligned}
\bm R_k(f)&=\frac{1}{J}\sum_{j=1}^{J}
\bm x_{k,j}(f)\bm x_{k,j}^{H}(f),\\
P_k(f,\theta)&=\left[
\bm a^{H}(f,\theta)\widetilde{\bm R}_k^{-1}(f)
\bm a(f,\theta)\right]^{-1}.
\end{aligned}
\label{eq:mvdr}
\end{equation}
Let \(p_k(\theta)\) and \(\bar p(\theta)\) denote the window-level and recording-level log-angular profiles. They are obtained by averaging over the analysis band and observation windows, respectively:
\begin{equation}
\begin{aligned}
p_k(\theta)&=\operatorname{mean}_{f}\log P_k(f,\theta),\\
\bar p(\theta)&=\operatorname{mean}_{k}p_k(\theta).
\end{aligned}
\label{eq:angular_profiles}
\end{equation}
\nopagebreak[4]
Evaluated over the bearing grid, their batched representations are \(\mathbf p_k,\bar{\mathbf p}\in\mathbb R^{B\times N_{\theta}}\). Stacking the \(K\) window profiles gives a tensor of size \(B\times K\times N_{\theta}\).

Circular convolutions encode the angular profiles. Their outputs are combined with angular-profile values and sine/cosine bearing encodings. Let \(\mathbf D_{\mathrm{win}}\) and \(\mathbf D_{\mathrm{raw}}\) denote the resulting window-level and recording-level directional features:
\begin{equation}
\begin{aligned}
\mathbf D_{\mathrm{win}}&\in\mathbb R^{B\times K\times N_{\theta}\times d},\\
\mathbf D_{\mathrm{raw}}&\in\mathbb R^{B\times N_{\theta}\times d}.
\end{aligned}
\label{eq:directional_features}
\end{equation}
These features guide the DET and LINE branches, respectively. Groups of 12 adjacent bearings in \(\mathbf D_{\mathrm{raw}}\) are flattened and passed through a shared linear map to obtain the directional sequence \(\mathbf D\in\mathbb R^{B\times30\times d_{\mathrm o}}\).

\subsubsection{Target detection branch DET}
\label{sec:content_branches}
\label{sec:tf_branch}

The DET branch extracts the real part, imaginary part, and log magnitude of each sensor's short-time Fourier spectrum. Time-frequency pooling, relative sensor-coordinate encoding, and attention-based sensor aggregation produce a grid with 8 time positions and 64 frequency positions per observation window. Parallel temporal and frequency Transformers model the two axes. Their outputs are added to the input grid, followed by layer normalization. Flattening the time-frequency axes gives the content tensor \(\mathbf H_{\mathrm{TF}}\):
\begin{equation}
\mathbf H_{\mathrm{TF}}\in\mathbb R^{B\times K\times512\times d},
\qquad 512=8\times64.
\label{eq:tf_content}
\end{equation}

Let \(\operatorname{MHA}\) denote multihead attention, with its three arguments ordered as query, key, and value inputs~\cite{uad_vaswani2017}. Within each observation window, the directional features provide the queries and the time-frequency features provide the keys and values. The attention output \(\mathbf A\) is
\begin{equation}
\begin{aligned}
\mathbf A&=\operatorname{MHA}\bigl(
\mathbf D_{\mathrm{win}},\mathbf H_{\mathrm{TF}},\mathbf H_{\mathrm{TF}}\bigr),\\
\mathbf A&\in\mathbb R^{B\times K\times N_{\theta}\times d}.
\end{aligned}
\label{eq:direction_time_attention}
\end{equation}
A gated residual connection and layer normalization integrate \(\mathbf A\) into the window-level directional features. Averaging over bearings and applying a shared linear map gives the temporal sequence \(\mathbf T_{\mathrm{ac}}\), retaining one representation per observation window:
\begin{equation}
\begin{aligned}
{}[B,K,N_{\theta},d]&\xrightarrow{\text{bearing mean}}[B,K,d],\\
[B,K,d]&\xrightarrow{\text{linear}}\mathbf T_{\mathrm{ac}}
\in\mathbb R^{B\times K\times d_{\mathrm o}}.
\end{aligned}
\label{eq:window_time_features}
\end{equation}

\subsubsection{Spectral-line detection branch LINE}
\label{sec:ff_branch}

The LINE branch extracts log power spectral density (PSD), local spectral prominence, and intersensor magnitude coherence from the fine spectra within each observation window. Local prominence measures the log-PSD excess over a smoothed frequency neighborhood. Magnitude coherence is derived from normalized cross-spectra averaged over selected sensor pairs. The mean and standard deviation of each feature across the \(K\) observation windows form six frequency channels. A multiscale frequency CNN encodes these channels into the frequency representation \(\mathbf F_0\):
\begin{equation}
\begin{aligned}
{}[B,K,N_f,3]
&\xrightarrow{\text{mean and std. over }K}
[B,N_f,6]\\
&\xrightarrow{\text{multiscale frequency CNN}}
\mathbf F_0\in\mathbb R^{B\times N_f\times d}.
\end{aligned}
\label{eq:fine_frequency_features}
\end{equation}

Frequency features query the recording-level directional features. Let \(\zeta_F\) be a learnable scalar gain and \(\mathbf F_{\mathrm{dir}}\) the frequency representation after directional fusion. Then
\begin{equation}
\begin{aligned}
\mathbf F_{\mathrm{dir}}={}&\mathbf F_0+
\zeta_F\operatorname{MHA}\bigl(
\mathbf F_0,\mathbf D_{\mathrm{raw}},
\mathbf D_{\mathrm{raw}}\bigr),\\
\mathbf F_{\mathrm{dir}}&\in
\mathbb R^{B\times N_f\times d}.
\end{aligned}
\label{eq:direction_frequency_attention}
\end{equation}

The frequency axis is zero-padded to a multiple of 32. Groups of 32 adjacent positions are then flattened and passed through a shared linear map, producing
\(\mathbf F\in\mathbb R^{B\times63\times d_{\mathrm o}}\).

\subsection{Task-Specific Acoustic Representation Learning}
\label{sec:task_representation}

Ordered flattening and independent linear maps convert \(\mathbf T_{\mathrm{ac}}\), \(\mathbf D\), and \(\mathbf F\) into task summaries \(\mathbf c_{\mathrm{det}}\), \(\mathbf c_{\mathrm{doa}}\), and \(\mathbf c_{\mathrm{freq}}\), respectively. Each task summary is supervised through its own prediction head. Table~\ref{tab:task_routing} gives the task correspondence.

\begin{table*}[t]
\centering
\caption{Acoustic sequences, task summaries, and supervision.}
\label{tab:task_routing}
\small
\setlength{\tabcolsep}{4.5pt}
\renewcommand{\arraystretch}{1.2}
\begin{tabularx}{\textwidth}{@{}lcc>{\raggedright\arraybackslash}X>{\raggedright\arraybackslash}X@{}}
\toprule
Task & Sequence & Summary & Prediction & Supervision\\
\midrule
Detection & \(\mathbf T_{\mathrm{ac}}\) & \(\mathbf c_{\mathrm{det}}\) & Target-presence probability & Binary cross-entropy\\
Bearing & \(\mathbf D\) & \(\mathbf c_{\mathrm{doa}}\) & Distribution over bearings & KL divergence with circular soft labels\\
Frequency & \(\mathbf F\) & \(\mathbf c_{\mathrm{freq}}\) & Independent scores over frequency bins & Region-balanced binary cross-entropy with hard negatives\\
\bottomrule
\end{tabularx}
\end{table*}

\noindent\textbf{Detection supervision.}
Let \(y_{\mathrm{det}}\in\{0,1\}\) indicate target presence and \(\widehat y_{\mathrm{det}}\in(0,1)\) be the detection head's sigmoid output. For one observation, the binary cross-entropy is
\begin{equation}
\begin{aligned}
\mathcal L_{\mathrm{det}}={}&-y_{\mathrm{det}}\log\widehat y_{\mathrm{det}}\\
&-(1-y_{\mathrm{det}})\log(1-\widehat y_{\mathrm{det}}).
\end{aligned}
\label{eq:detection_objective}
\end{equation}
Training averages this loss over the batch.

\noindent\textbf{Bearing supervision.}
Let \(q_{\theta}\) and \(\widehat q_{\theta}\) be the reference and predicted bearing probabilities. The reference is a normalized Gaussian soft label based on the shortest circular distance from the true bearing; the prediction is a softmax distribution. For a target-present observation,
\begin{equation}
\mathcal L_{\mathrm{doa}}=
\sum_{\theta}q_{\theta}
\log\frac{q_{\theta}}{\widehat q_{\theta}}.
\label{eq:doa_objective}
\end{equation}
The sum covers the candidate bearings, and training averages the loss over target-present observations.

\noindent\textbf{Frequency supervision.}
Each reference line forms a local Gaussian soft peak on the fine-frequency grid; overlapping peaks are combined by their maximum. All frequency bins are supervised, with all-zero targets for target-absent observations. Let \(y\in[0,1]\) and \(\widehat y\in(0,1)\) denote a bin's soft target and independent sigmoid prediction, and let \(\ell\) denote its binary cross-entropy. For one observation, \(\bar\ell_+\) and \(\bar\ell_-\) are the mean losses over positive and negative bins, while \(\bar\ell_{\mathrm h}\) is the mean over the highest-scoring negative bins. With nonnegative weights \(\beta_-\) and \(\beta_h\),
\begin{equation}
\begin{aligned}
\ell(y,\widehat y)&=-y\log\widehat y-(1-y)\log(1-\widehat y),\\
\mathcal L_{\mathrm{freq}}&=\bar\ell_+
+\beta_-\bar\ell_-+\beta_h\bar\ell_{\mathrm h}.
\end{aligned}
\label{eq:frequency_objective}
\end{equation}
Separate means balance sparse line targets and numerous negative bins. The hard-negative term adds weight to high-scoring negative bins. Target-absent observations use the two negative-bin terms. Sample losses are averaged over the batch. Section~\ref{sec:setup} gives the label and loss settings.

Nonnegative coefficients \(\lambda_{\mathrm{det}}\), \(\lambda_{\mathrm{doa}}\), and \(\lambda_{\mathrm{freq}}\) balance the task losses:
\begin{equation}
\mathcal L_{\mathrm{ac}}=
\lambda_{\mathrm{det}}\mathcal L_{\mathrm{det}}
+\lambda_{\mathrm{doa}}\mathcal L_{\mathrm{doa}}
+\lambda_{\mathrm{freq}}\mathcal L_{\mathrm{freq}}.
\label{eq:acoustic_objective}
\end{equation}
These losses train the task summaries and the branch and fusion modules that produce them.

\subsection{Multimodal Alignment and Parameter-Efficient Adaptation}
\label{sec:multimodal_alignment}

\noindent\textbf{Acoustic and environmental inputs.}
Independent linear maps project the three task summaries and the \(\mathbf D\), \(\mathbf T_{\mathrm{ac}}\), and \(\mathbf F\) sequences into the language-model embedding space. Each task summary occupies one input position. Group embeddings identify the source of local acoustic tokens, and within-group positional encodings retain their order.

Bathymetry is supplied as a numerical depth matrix. After spatial pooling and normalization, local depth patches and their center coordinates are mapped to environmental tokens by a multilayer perceptron. Task instructions, sound speed, receiver coordinates, and array geometry are encoded as text. One additional frequency-readout position, derived from \(\mathbf c_{\mathrm{freq}}\), is appended after the multimodal inputs. Its causal attention can access the complete input context. The structured answer follows this position.

\noindent\textbf{Two-stage adaptation.}
The first stage learns the acoustic encoder and task summaries with Eq.~\eqref{eq:acoustic_objective}. The second stage freezes the acoustic encoder and base language-model parameters while training the input projections, bathymetric encoder, low-rank adaptation (LoRA) modules, and final frequency readout. LoRA is applied to the language model's query and value projections~\cite{uad_hu2022lora}. Section~\ref{sec:setup} specifies the task sequence.

\noindent\textbf{Language and frequency supervision.}
Target presence, bearing, received SNR, and horizontal range are supervised through their field values in the reference answer. Let \(\mathcal A\) collect answer-token positions across the batch, excluding input tokens. Partition it into disjoint subsets \(\mathcal A_u\), with \(u\in\{\mathrm{det},\mathrm{doa},\mathrm{snr},\mathrm{range},\mathrm{rest}\}\). The remaining-text group includes field names, structural tokens, and explanatory text. Let \(\pi_i\) be the probability of the reference token at position \(i\), conditioned on the multimodal input and preceding reference tokens. The contribution of group \(u\) is
\begin{equation}
\mathcal L_u^{\mathrm{text}}=
-\frac{1}{|\mathcal A|}\sum_{i\in\mathcal A_u}\log\pi_i.
\label{eq:field_text_objective}
\end{equation}
The shared denominator makes the five contributions sum to the mean answer-token cross-entropy. For target-absent observations, bearing, SNR, and range are supervised as \texttt{null}.

The final frequency head reads the language-model hidden state at the appended readout position. Its loss \(\mathcal L_{\mathrm{freq}}^{\mathrm{LLM}}\) uses the same soft targets, weights, and hard-negative selection rule as the encoder frequency loss in Eq.~\eqref{eq:frequency_objective}. Each head selects hard negatives from its own predicted scores. With nonnegative frequency-loss weight \(\eta\), the alignment objective is
\begin{equation}
\begin{aligned}
\mathcal L_{\mathrm{align}}={}&
\mathcal L_{\mathrm{det}}^{\mathrm{text}}+
\mathcal L_{\mathrm{doa}}^{\mathrm{text}}+
\mathcal L_{\mathrm{snr}}^{\mathrm{text}}+
\mathcal L_{\mathrm{range}}^{\mathrm{text}}\\
&+\mathcal L_{\mathrm{rest}}^{\mathrm{text}}+
\eta\mathcal L_{\mathrm{freq}}^{\mathrm{LLM}}.
\end{aligned}
\label{eq:alignment_objective}
\end{equation}
Each adaptation step uses the supervision assigned to its tasks.

\noindent\textbf{Structured output.}
The language model generates target presence, bearing, horizontal range, and received SNR. The frequency head produces independent sigmoid scores. Thresholding, local-peak selection, and minimum-separation suppression yield a variable-length line list, with each frequency paired with its sigmoid score. The scalar fields, frequency list, and explanatory text form the structured report. Target-absent reports use an empty frequency list and null physical attributes.

\section{Physics-Parameterized Data Engine for Passive Detection}
\label{sec:data}

The data engine combines source synthesis, array propagation, and controlled noise mixing. Each array observation is paired with its source composition, geometry, propagation environment, and target attributes. White noise and a band-extended measured background yield two datasets for learning acoustic representations and structured detection descriptions.

\subsection{Source Signal Synthesis}
\label{sec:source}

The source model combines tonal, continuous-spectrum, and modulation-spectrum components. Let $t$ denote time and $s_{\mathrm{src}}(t)$ the source signal. The tonal, modulation, and continuous-spectrum components are denoted by $s_{\mathrm{tonal}}(t)$, $s_{\mathrm{mod}}(t)$, and $s_{\mathrm{broad}}(t)$, respectively. With their amplitude weights $w_{\mathrm{tonal}}$, $w_{\mathrm{mod}}$, and $w_{\mathrm{broad}}$, the source is
\begin{equation}
\begin{aligned}
s_{\mathrm{src}}(t)
={}&w_{\mathrm{tonal}}s_{\mathrm{tonal}}(t)
+w_{\mathrm{mod}}s_{\mathrm{mod}}(t)\\
&+w_{\mathrm{broad}}s_{\mathrm{broad}}(t).
\end{aligned}
\label{eq:source_total}
\end{equation}
The weights absorb the root-mean-square (RMS) normalization of the individual components. Frequency, amplitude, initial phase, and modulation parameters vary across source realizations.

\subsubsection{Tonal component}
The tonal component contains selected in-band harmonics of a fundamental frequency $f_0$ and independently sampled spectral lines. A harmonic carrier has frequency $qf_0$, where $q$ is its integer order. Index all retained carriers by $\ell=1,\ldots,L_{\mathrm{src}}$, where $L_{\mathrm{src}}$ is the number of source lines. For line $\ell$, let $f_\ell$, $A_\ell$, $\varphi_\ell$, and $\dot f_\ell$ denote its initial frequency, amplitude, initial phase, and constant frequency-drift rate, respectively. The tonal signal is
\begin{equation}
s_{\mathrm{tonal}}(t)
=\sum_{\ell=1}^{L_{\mathrm{src}}}A_\ell
\cos\!\left[2\pi\!\left(f_\ell t+\tfrac12\dot f_\ell t^2\right)+\varphi_\ell\right].
\label{eq:tonal}
\end{equation}
Frequencies are measured in Hz and drift rates in Hz/s. Harmonic amplitudes decrease with order; the amplitudes of independent lines are sampled separately.

\subsubsection{Continuous-spectrum component}
The continuous spectrum is generated by an autoregressive (AR) model driven by zero-mean Gaussian white noise. Let $p_{\mathrm{AR}}$ be the model order, $b_\ell$ its coefficients, and $\epsilon[n]$ the excitation at sample $n$. The sampled component satisfies
\begin{equation}
s_{\mathrm{broad}}[n]
=\sum_{\ell=1}^{p_{\mathrm{AR}}}b_\ell s_{\mathrm{broad}}[n-\ell]+\epsilon[n].
\label{eq:broadband}
\end{equation}
A prescribed autocorrelation sequence is constructed as an exponentially damped cosine, with its oscillation rate and decay controlling spectral center and bandwidth. The AR coefficients are obtained from the Yule--Walker equations.

\subsubsection{Modulation-spectrum component}
Let $g_{\mathrm{mod}}(t)$ be a train of equally spaced Gaussian pulses. Periodic amplitude fluctuations of the continuous spectrum are represented by
\begin{equation}
s_{\mathrm{mod}}(t)=g_{\mathrm{mod}}(t)\,s_{\mathrm{broad}}(t).
\label{eq:modulated}
\end{equation}
Pulse period and width determine the repetition rate and duration of the amplitude fluctuations.

\subsection{Propagation and Background Noise}
\label{sec:channel}

For sensor $m$, let $h_m(t)$ denote the channel impulse response and $s_{\mathrm p,m}(t)$ the propagated target signal. With $*$ denoting convolution,
\begin{equation}
s_{\mathrm p,m}(t)=(h_m*s_{\mathrm{src}})(t),
\qquad m=0,\ldots,M_{\mathrm a}-1.
\label{eq:array_propagation}
\end{equation}
The channels share a common time reference, preserving relative propagation delays across sensors.

\subsubsection{Environment and array geometry}
Horizontal positions use Cartesian coordinates $x$ and $y$, with depth positive downward from the sea surface. The simulated bathymetry covers $[-50,50]\times[-50,50]$~km. A cosine slope transitions from a depth of $3$~km at the left boundary to $1$~km at the right boundary, with smooth lateral variations of at most $100$~m. Water depth at the array center is $1.9$~km.

\begin{figure*}[!t]
\centering
\includegraphics[width=0.95\textwidth]{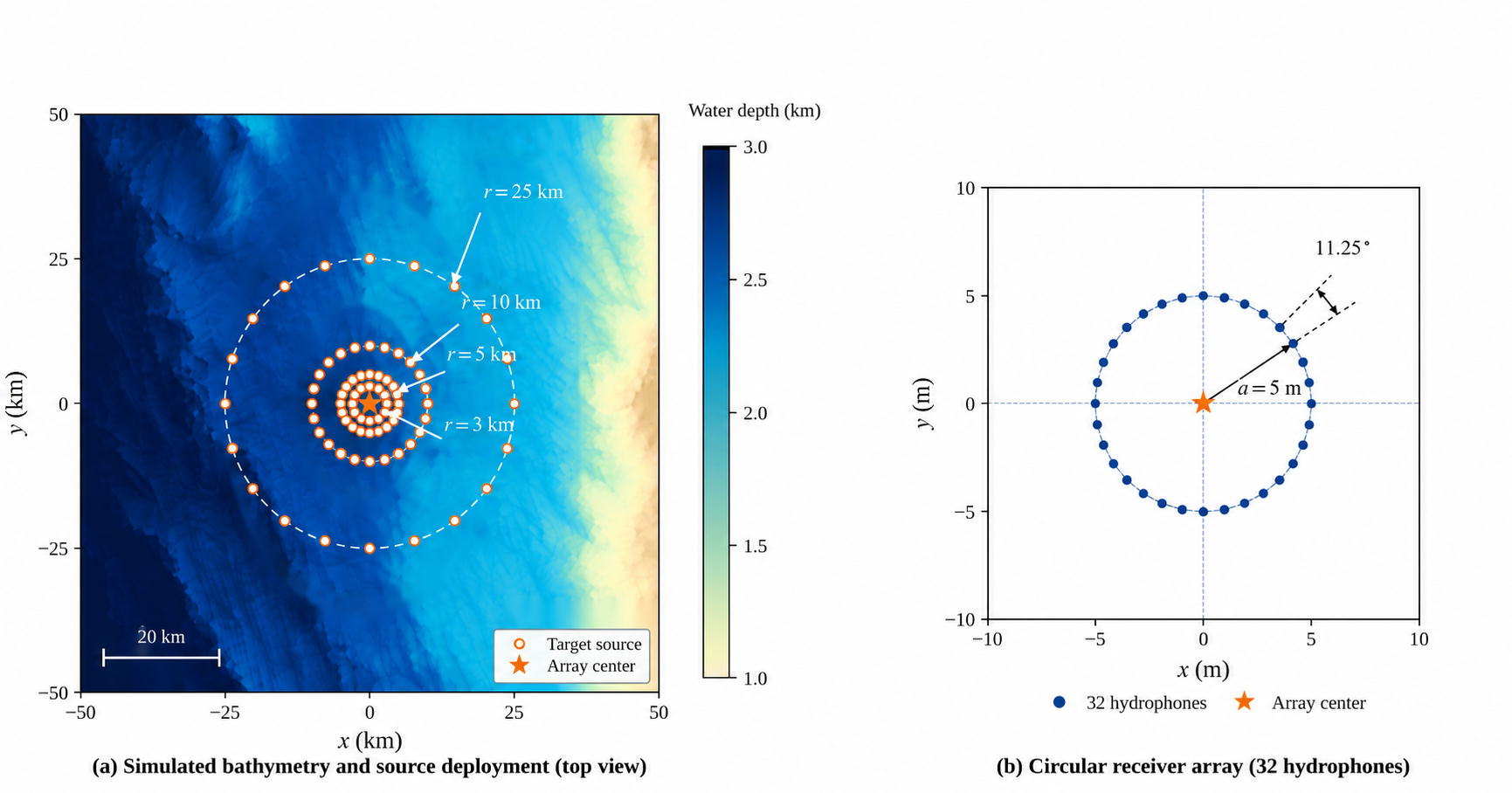}
\caption{Propagation environment and array geometry. (a) Simulated Bathymetry over the horizontal domain, with water depth in km. Open circles mark candidate target locations and the star marks the array center. (b) Receiver-array plan view, with coordinates in m. Filled circles indicate hydrophones and $a$ denotes array radius.}
\label{fig:terrain_array}
\end{figure*}

The receiver is a horizontal uniform circular array with $32$ hydrophones, radius $a=5$~m, and angular spacing $11.25^\circ$. Its horizontal center is the origin. Target bearing $\theta$ is measured counterclockwise from the positive $x$ axis, and $r$ is horizontal range from the array center. Sources occupy four concentric circles with radii $3$, $5$, $10$, and $25$~km, corresponding to diameters $6$, $10$, $20$, and $50$~km. Each circle has $24$ uniformly spaced positions at $15^\circ$ intervals, yielding $96$ geometries. Source and receiver depths are both $100$~m. Figure~\ref{fig:terrain_array} shows the layout; each source position generates separate single-target observations.

\begin{table}[!t]
\centering
\caption{Propagation Configuration}
\label{tab:environment}
\footnotesize
\begin{tabularx}{\columnwidth}{@{}p{0.31\columnwidth}X@{}}
\toprule
Parameter & Setting \\
\midrule
Bathymetric grid & $201\times201$ points; $0.5$~km spacing \\
Water sound speed & Constant, $1500$~m/s \\
Seabed properties & Sound speed $1700$~m/s; density $1.8$~g/cm$^3$; attenuation $0.5$~dB/wavelength \\
Propagation solver & $30$--$830$~Hz at $40$~Hz spacing; $8000$ beams; launch angles $\pm60^\circ$ \\
\bottomrule
\end{tabularx}
\end{table}

BELLHOP~\cite{uad_porter2011bellhop} computes propagation along a radial vertical slice extracted from the global bathymetric grid between the source and the array center. The actual source-to-sensor distances determine the receiver ranges. Responses at the frequency nodes in Table~\ref{tab:environment} are interpolated across the source band. A common RMS normalization over the propagated multichannel record preserves relative sensor amplitudes and phases.

\subsubsection{Background construction and SNR}
The white background is Gaussian noise independent across time and sensors. The second background combines synchronized low-frequency ocean recordings with synthetic high-frequency noise. For each observation, $32$ of the $70$ recorded channels are selected without replacement. They share the same crop interval, and the channel mapping remains fixed throughout the record. The recordings are treated as $500$~Hz data and resampled to $2$~kHz. The retained band through $250$~Hz is combined with independently generated noise from $250$ to $1000$~Hz. Its power is scaled per channel to match the mean low-band power spectral density, excluding the DC component. We refer to this construction as the \emph{mixed background}.

Let $\mathbf s_{\mathrm p}(t)$ collect the propagated sensor signals, $\mathbf n(t)$ denote the background vector, and $\mathbf x(t)$ the received observation. Let $P_s$ and $P_n$ be the respective mean powers over all sensors and the complete observation. For a prescribed broadband received SNR $\gamma$ in dB, the mixture and target amplitude factor $\alpha$ are
\begin{equation}
\mathbf x(t)=\alpha\mathbf s_{\mathrm p}(t)+\mathbf n(t),
\qquad
\alpha=\sqrt{10^{\gamma/10}\frac{P_n}{P_s}}.
\label{eq:snr_mixing}
\end{equation}
Background-only observations omit the injected target and reuse the same background draw as their target-present counterparts. All sensors then receive one common RMS scaling. Sampling and batching these observations gives the encoder input $\mathbf X$.

\subsection{Dual-Granularity Supervision}
\label{sec:supervision}

\noindent\textbf{Physics-aware acoustic representation labels (PAR).}
The array waveform $\mathbf X$ is paired with target-presence, bearing, and line-location labels. The binary label $y_{\mathrm{det}}$ specifies whether a target is injected. Bearing supervision uses a circular soft distribution on a $1^\circ$ grid for target-present samples. Frequency labels cover the full $0.5$~Hz grid over $0$--$1000$~Hz. Reference target lines are identified from received spectral peaks, the propagated clean signal, and source parameters. These reference lines are converted into the soft targets defined in Section~\ref{sec:task_representation}. Background-only samples have $y_{\mathrm{det}}=0$ and all-zero frequency targets.

\noindent\textbf{Structured acoustic descriptions (SAD).}
Continuous acoustic features and task instructions form the input to the language interface. Range and SNR tasks additionally receive the global bathymetric grid, receiver location, array geometry, and sound-speed parameters. The structured targets are \texttt{target\_present} ($y_{\mathrm{det}}$), \texttt{doa\_deg} ($\theta$, degrees), \texttt{range\_m} ($r$, m), \texttt{snr\_db} ($\gamma$, dB), and \texttt{frequencies\_hz} (Hz). Presence, bearing, range, and SNR use answer-token supervision; line locations supervise the multilabel readout from language hidden states. Target coordinates, injected SNR, and reference spectral lines appear only in the supervision.

\subsection{Dataset Configuration and Partitioning}
\label{sec:datastats}

Each observation contains $120$~s of synchronized data sampled at $2$~kHz, yielding $N=240{,}000$ samples per sensor. Eight source realizations are generated at each of the $96$ geometries, producing $768$ source instances. The two background datasets share the source and propagation configurations. Target-present observations use SNRs of $0$, $-10$, $-20$, $-25$, and $-30$~dB.

The eight realizations at each geometry are assigned in a $5{:}1{:}1{:}1$ ratio to training (TRAIN), development (DEV), calibration (CAL), and test (TEST) partitions. All observations derived from one source instance remain in the same partition. Source realizations and clean waveforms are disjoint across partitions; the same geometry grid and environment are represented in each. Measured background data are divided into nonoverlapping temporal blocks separated by unused intervals. This partition evaluates new source realizations within the specified environment and geometry grid.

TRAIN uses one background draw per source, with five target-present SNR conditions and one background-only observation. DEV pairs a $-30$~dB observation with its background. TEST uses three background draws per source, each producing all five SNR conditions and one background-only observation. CAL contains $288$ background-only observations per background for detector calibration. Table~\ref{tab:partitions} reports the resulting observation counts. DEV supports model selection. The UPD decision rules are specified in Section~\ref{sec:metrics}.

\begin{table}[!t]
\centering
\caption{Observation Counts for Each Background Dataset}
\label{tab:partitions}
\footnotesize
\setlength{\tabcolsep}{3pt}
\begin{tabular*}{\columnwidth}{@{\extracolsep{\fill}}lrrrr@{}}
\toprule
Partition & Source IDs & Target & Background & Total \\
\midrule
TRAIN & 480 & 2400 & 480 & 2880 \\
DEV & 96 & 96 & 96 & 192 \\
CAL & 96 & 0 & 288 & 288 \\
TEST & 96 & 1440 & 288 & 1728 \\
\bottomrule
\end{tabular*}
\par\smallskip
\begin{minipage}{\columnwidth}\footnotesize
The counts apply separately to white and mixed backgrounds. Background observations are counted once per draw, irrespective of the number of SNR conditions.
\end{minipage}
\end{table}

\section{Experiments}
\label{sec:experiments}

This section specifies the implementation and evaluates target presence, bearing, spectral lines, received SNR, and horizontal range under white noise and a mixed background.

\subsection{Experimental Setup}
\label{sec:setup}

\subsubsection{Implementation Details}
The branches use the shared observation windows defined in Section~\ref{sec:direction_guided_encoding}. Table~\ref{tab:fft} lists their internal Fourier settings. All transforms use periodic Hann windows. Each DOA observation window contains $J=39$ spectral snapshots. The bearing grid has $N_\theta=360$ positions, from $0^\circ$ to $359^\circ$ in $1^\circ$ steps. The LINE grid contains $N_f=2001$ frequency bins over 0--1{,}000~Hz.

\begin{table}[!t]
\centering
\caption{Fourier analysis within the shared observation windows.}
\label{tab:fft}
\footnotesize
\renewcommand{\arraystretch}{1.15}
\setlength{\tabcolsep}{3pt}
\begin{tabular*}{\columnwidth}{@{\extracolsep{\fill}}lrrrrl@{}}
\toprule
Branch & \shortstack{Analysis\\window (s)} & \shortstack{Analysis\\hop (s)} & \shortstack{FFT\\points} & \shortstack{Bin spacing\\(Hz)} & \shortstack{Band\\(Hz)} \\
\midrule
DOA & 0.5 & 0.25 & 1{,}000 & 2.0 & 40--700 \\
DET & 0.5 & 0.25 & 1{,}000 & 2.0 & 0--1{,}000 \\
LINE & 2.0 & 1.00 & 4{,}000 & 0.5 & 0--1{,}000 \\
\bottomrule
\end{tabular*}
\end{table}

The internal and output feature widths are $d=64$ and $d_{\mathrm o}=128$. Excluding the batch axis, $\mathbf D$, $\mathbf T_{\mathrm{ac}}$, and $\mathbf F$ have shapes $30\times128$, $56\times128$, and $63\times128$, respectively. Ordered flattening and independent linear projections form detection, bearing, and frequency summaries of 128, 128, and 256 dimensions. The 149 local tokens and three task summaries occupy 152 acoustic-content positions.

The language backbone is Qwen3-4B-Instruct-2507~\cite{uad_qwen3_4b_instruct2507,uad_qwen3technicalreport2025}, with hidden width 2{,}560. LoRA uses rank 8, an alpha parameter of 16, and dropout 0.05 in the query and value projections. The $201\times201$ bathymetric grid is mapped to 16 tokens; receiver metadata is supplied as text. Including the additional frequency-readout position, the multimodal prefix contains 169 nontext positions plus instruction and metadata tokens. The answer follows the readout position.

Computation uses NVIDIA RTX~3090 GPUs with 24~GB memory. Spectral and spatial statistics use FP32, and language adaptation uses BF16 mixed precision. Optimization uses AdamW~\cite{uad_loshchilov2019adamw}, weight decay 0.01, an effective batch size of 8, and gradient norm clipping at 1.0.

Diagonal loading adds 0.03 times the mean diagonal power to each covariance diagonal entry. Bearing soft labels have a Gaussian standard deviation of $5^\circ$. Frequency soft peaks have a standard deviation of 1~Hz and are truncated at $\pm2$~Hz. Bins with soft targets above 0.01 are positive; the remaining bins are negative. The frequency loss uses $\beta_-=2$ and $\beta_h=0.5$. For each observation, the hard-negative term averages the losses of the 32 highest-scoring negative bins, or all negative bins when fewer are available. These settings apply to both frequency heads.

\begin{table*}[!t]
\centering
\caption{Encoder validation performance at $-30$~dB.}
\label{tab:frontend}
\small
\renewcommand{\arraystretch}{1.15}
\setlength{\tabcolsep}{3pt}
\begin{tabular*}{\textwidth}{@{\extracolsep{\fill}}lrrrrrr@{}}
\toprule
Background & $P_{\mathrm d}$ (\%) & \shortstack{False alarms\\$N_{\mathrm{FP}}/N_0$} & \shortstack{Bearing\\MAE ($^\circ$)} & Precision & Recall & F1 \\
\midrule
White noise & 100.00 & 0/96 & 1.26 & 100.00 & 93.64 & 96.72 \\
Mixed background & 98.96 & 0/96 & 0.22 & 81.75 & 81.83 & 81.79 \\
\bottomrule
\end{tabular*}
\par\vspace{4pt}
\noindent\parbox{\textwidth}{\footnotesize Each row uses a separately trained encoder and 96 target-present plus 96 target-absent validation observations. Bearing MAE includes all target-present observations. Frequency Precision, Recall, and F1 follow Section~\ref{sec:metrics} and are percentages.}
\end{table*}

\begin{table*}[!t]
\centering
\caption{System test performance under the mixed background.}
\label{tab:system_results}
\small
\renewcommand{\arraystretch}{1.18}
\setlength{\tabcolsep}{5pt}
\begin{tabular*}{\textwidth}{@{\extracolsep{\fill}}rrrrrrrrr@{}}
\toprule
\shortstack{SNR (dB)} & \shortstack{$P_{\mathrm d}$ (\%)} & \shortstack{$P_{\mathrm{fa}}$ (\%)} & \shortstack{Bearing MAE ($^\circ$)} & \shortstack{Precision (\%)} & \shortstack{Recall (\%)} & \shortstack{F1 (\%)} & \shortstack{SNR MAE (dB)} & \shortstack{Range MAE (km)} \\
\midrule
0     & 100.00 & -- & 0.000 & 77.54 & 35.74 & 48.93 & 0.000 & 8.410 \\
$-10$ & 100.00 & -- & 0.000 & 65.74 & 38.25 & 48.36 & 1.979 & 8.222 \\
$-20$ & 100.00 & -- & 0.260 & 65.76 & 45.34 & 53.68 & 0.451 & 9.174 \\
$-25$ & 100.00 & -- & 0.260 & 64.81 & 51.84 & 57.61 & 0.625 & 9.462 \\
$-30$ & 100.00 & -- & 0.521 & 63.20 & 56.78 & 59.82 & 0.105 & 8.749 \\
\midrule
Overall & 100.00 & 0.00 & 0.208 & 67.11 & 43.95 & 53.11 & 0.632 & 8.803 \\
\bottomrule
\end{tabular*}
\par\vspace{4pt}
\noindent\parbox{\textwidth}{\footnotesize Frequency Precision, Recall, and F1 follow Section~\ref{sec:metrics} and are percentages. Each SNR row contains 288 target-present observations. False alarms are evaluated once on 288 shared target-absent observations, giving 0/288, and are shown only in the overall row. SNR and range have numerical outputs for 287/288 observations at $-30$~dB and 288/288 at every other SNR, giving 1{,}439/1{,}440 overall.}
\end{table*}

\begin{figure*}[!t]
\centering
\includegraphics[width=\textwidth]{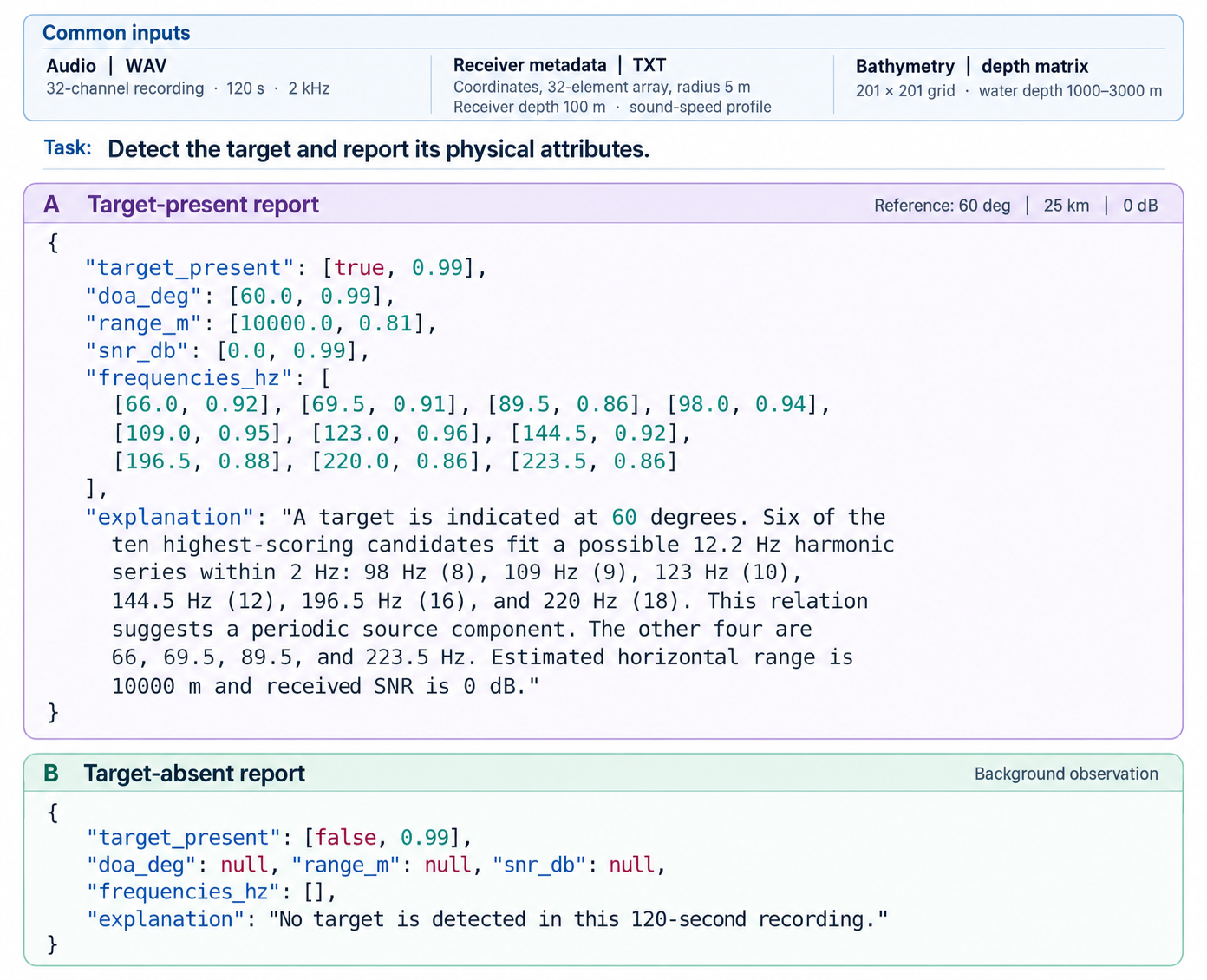}
\caption{Receiver inputs and structured-report examples. Both observations use the same receiver metadata and numerical bathymetry. The target-present case has reference bearing $60^\circ$, horizontal range 25~km, and SNR 0~dB. Numerical fields and all 23 frequency candidates are saved predictions. The English explanation is an edited illustration of how a possible harmonic relation can be reported. The target-absent case returns null physical attributes and an empty frequency list. Support scores are truncated to two decimal places.}
\label{fig:io}
\end{figure*}

\subsubsection{Training Protocol}
After acoustic pretraining, the encoder is frozen. Language adaptation proceeds sequentially: one epoch for detection and bearing, two epochs for frequency prediction, and one epoch for SNR, horizontal range, and explanation text. Environmental text and bathymetry are added in the last step.

\subsection{Evaluation Protocol}
\label{sec:metrics}

Data partitions follow Section~\ref{sec:data}. Encoder validation uses a separately trained model for each background, with 96 target-present observations at $-30$~dB and 96 target-absent observations. The test partition contains disjoint source realizations and was inspected during development. The reported evaluation uses fixed model parameters and decision rules. There are 288 target-present observations at each of 0, $-10$, $-20$, $-25$, and $-30$~dB, together with 288 shared target-absent observations. The background observations enter aggregate statistics once.

\subsubsection{Target Detection}
Let $N_1$ and $N_0$ denote the numbers of target-present and target-absent observations, respectively. Let $N_{\mathrm{TP}}$ be the number of correctly detected targets and $N_{\mathrm{FP}}$ the number of false alarms. The empirical detection and false-alarm rates are
\begin{equation}
P_{\mathrm d}=\frac{N_{\mathrm{TP}}}{N_1},\qquad
P_{\mathrm{fa}}=\frac{N_{\mathrm{FP}}}{N_0}.
\label{eq:detection_metrics}
\end{equation}
The encoder declares a target when its presence probability exceeds 0.5. The system output is scored from the generated \texttt{target\_present} field.

\subsubsection{Bearing Estimation}
For target-present observation $i$, let $\theta_i$ and $\widehat\theta_i$ denote the reference and predicted bearings in $[0^\circ,360^\circ)$. The circular error and its mean absolute value are
\begin{equation}
\begin{aligned}
e_i&=\min\!\left(\left|\widehat\theta_i-\theta_i\right|,
360^\circ-\left|\widehat\theta_i-\theta_i\right|\right),\\
\mathrm{MAE}_{\theta}&=\frac{1}{N_1}\sum_{i=1}^{N_1}e_i.
\end{aligned}
\label{eq:doa_mae}
\end{equation}
All target-present observations contribute to this metric. The test set contains 24 reference bearings, $0^\circ,15^\circ,\ldots,345^\circ$.

\subsubsection{Frequency Outputs}
The language hidden-state readout assigns a score to each frequency bin. Local maxima with sigmoid scores of at least 0.5 are retained, followed by suppression that enforces a separation greater than 2~Hz.

For each observation, evaluated frequency candidates are processed in increasing frequency order and paired with the nearest unmatched reference within $\pm2$~Hz. Each candidate and reference can be used once. Let $L_{\mathrm{TP}}$, $L_{\mathrm{FP}}$, and $L_{\mathrm{FN}}$ denote the pooled counts of matched lines, unmatched evaluated candidates, and unmatched reference lines, respectively. The micro-averaged metrics are
\begin{equation}
\begin{aligned}
\mathrm{Precision}&=\frac{L_{\mathrm{TP}}}{L_{\mathrm{TP}}+L_{\mathrm{FP}}},\\
\mathrm{Recall}&=\frac{L_{\mathrm{TP}}}{L_{\mathrm{TP}}+L_{\mathrm{FN}}},\\
\mathrm{F1}&=\frac{2L_{\mathrm{TP}}}{2L_{\mathrm{TP}}+L_{\mathrm{FP}}+L_{\mathrm{FN}}}.
\end{aligned}
\label{eq:line_metrics}
\end{equation}

\subsubsection{SNR, Range, and Output Validity}
Received SNR and horizontal range are evaluated by MAE over target-present observations with valid numerical predictions. Numerical coverage is reported with each result. Schema validity checks the four scalar field types and the target-absent null rules defined in Section~\ref{sec:multimodal_alignment}.

\subsection{Acoustic Representation Performance}
\label{sec:representation}

Table~\ref{tab:frontend} reports encoder validation at $-30$~dB. The white-noise model detects 96 of 96 targets, and the mixed-background model detects 95 of 96. Both produce zero false alarms in 96 target-absent observations. Bearing MAE is $1.26^\circ$ and $0.22^\circ$, respectively. Frequency F1 is 96.72\% and 81.79\%. These results describe separately trained encoders evaluated on different background datasets.

\subsection{Detection and Structured Reports}
\label{sec:end_to_end}

\subsubsection{Physical Output Performance}
Table~\ref{tab:system_results} reports system evaluation under the mixed background. Target presence, bearing, SNR, and range are generated as structured fields; frequencies are read from the language hidden state.

The evaluated system detects all 1{,}440 targets and produces zero false alarms among 288 background observations. Overall bearing MAE is $0.208^\circ$, increasing to $0.521^\circ$ at $-30$~dB. Of the 1{,}440 bearing outputs, 1{,}420 equal the reference angle and 20 differ by $15^\circ$. These errors are measured on the 24 tested bearings.

Frequency precision, recall, and F1 are 67.11\%, 43.95\%, and 53.11\%, respectively. The system matches 10{,}927 of 24{,}862 reference lines. Reference-line counts decrease from 6{,}086 at 0~dB to 3{,}424 at $-30$~dB because annotation depends on the received spectrum. The increase in F1 at lower SNR therefore involves different reference sets.

Overall SNR MAE is 0.632~dB, with the largest error at $-10$~dB. Horizontal-range MAE is 8.803~km. A constant predictor equal to the training median of 7.5~km yields a lower MAE of 6.753~km on the same 1{,}439 observations. Per-record SNR scaling and common RMS normalization remove absolute received level as a unique range cue; any useful range information must remain in relative spatial or propagation structure.

\subsubsection{Structured-Report Examples}
The four scalar fields satisfy the joint type and null-value rules in 1{,}727 of 1{,}728 reports. One target-present observation at $-30$~dB has null SNR and range, while its detection, bearing, and frequency outputs remain available.

In Fig.~\ref{fig:io}, each displayed pair contains a value and its support score. Scalar scores are geometric means of the conditional probabilities of the tokens forming the reported field value. Frequency scores are the sigmoid outputs at the corresponding bins. Scores are truncated to two decimal places for display.

The positive example retains all 23 frequency candidates. Six of the ten highest-scoring candidates lie within $\pm2$~Hz of integer multiples of a possible 12.2~Hz fundamental, corresponding to a repetition period of approximately 82~ms. The illustrative explanation relates this pattern to a possible periodic source component and lists the other four candidates separately. The predicted bearing of $60^\circ$ agrees with the reference, whereas the predicted range of 10~km differs from the reference range of 25~km. The report lists the predicted attributes and the proposed spectral relation for inspection.


\section{Conclusion}
\label{sec:conclusion}

In this paper, we have developed a physics-guided audio-language approach for underwater passive target detection, building on the unified framework introduced in Part~I. The proposed acoustic encoder combines spatial, temporal, and spectral information from multichannel observations, using directional features to guide the extraction of weak acoustic structures. These representations are aligned with a pretrained language model and supplemented with environmental information to produce detection descriptions containing target-presence decisions and associated physical attributes. A physics-parameterized simulation pipeline provides multichannel recordings with traceable supervision, supporting both acoustic representation learning and language adaptation. Experiments with simulated targets under white and mixed backgrounds provide initial evidence for the feasibility of the proposed approach to passive acoustic detection. Together with the recognition study in Part~I, this work extends the framework toward a more complete account of underwater acoustic targets, from detecting their presence to describing their characteristics. 

Future work may examine the approach under changing propagation conditions, diverse sensor configurations, and continuous observations of moving targets, including validation with field recordings. It will also investigate how generated descriptions relate to the acoustic information underlying detection decisions.

	\bibliographystyle{ieeetr}
	\bibliography{ref}

\end{document}